\documentclass[conference]{IEEEtran}
\IEEEoverridecommandlockouts
\usepackage{cite}
\usepackage{amsmath,amssymb,amsfonts}
\usepackage{algorithmic}
\usepackage{graphicx}
\usepackage{textcomp}
\usepackage{xcolor}
\usepackage{booktabs}
\usepackage{caption}
\usepackage{hyperref}
\usepackage{colortbl}
\usepackage{amsmath}
 \usepackage{bbm}
\usepackage[normalem]{ulem}
\definecolor{lightgreen}{rgb}{0.7,1,0.9}
\def\BibTeX{{\rm B\kern-.05em{\sc i\kern-.025em b}\kern-.08em
    T\kern-.1667em\lower.7ex\hbox{E}\kern-.125emX}}
\begin{document}

\title{CutQAS: Topology-aware quantum circuit cutting via reinforcement learning}


\author{
\IEEEauthorblockN{
    Abhishek Sadhu\IEEEauthorrefmark{1}$^{,a}$,
    Aritra Sarkar\IEEEauthorrefmark{2}$^{,b}$, 
    Akash Kundu\IEEEauthorrefmark{3}$^{,c}$
}

\IEEEauthorblockA{\IEEEauthorrefmark{1} Centre for Quantum Science and Technology,\\ International Institute of Information Technology, Hyderabad, Telangana, India}
    
\IEEEauthorblockA{\IEEEauthorrefmark{1}\IEEEauthorrefmark{2}\IEEEauthorrefmark{3} Quantum Intelligence Alliance, Kolkata, India}

\IEEEauthorblockA{\IEEEauthorrefmark{2}{QWorld Association, Tallinn, Estonia}}

\IEEEauthorblockA{\IEEEauthorrefmark{3} QTF Centre of Excellence, Department of Physics, University of Helsinki, Finland}

\IEEEauthorblockA{$^a$\small:
\href{mailto:sadhuabhishek1@gmail.com}{sadhuabhishek1@gmail.com} (\small {corresponding author})}

\IEEEauthorblockA{$^b$\small: \href{mailto:aritra.sarkar@qworld.net}{aritra.sarkar@qworld.net}}

\IEEEauthorblockA{$^c$\small: \href{mailto:akash.kundu@helsinki.fi}{akash.kundu@helsinki.fi}}

}

\maketitle

\thispagestyle{plain}
\pagestyle{plain}

\begin{abstract}
Simulating molecular systems on quantum processors has the potential to surpass classical methods in computational resource efficiency.
The limited qubit connectivity, small processor size, and short coherence times of near-term quantum hardware constrain the applicability of quantum algorithms like QPE and VQE. 
Quantum circuit cutting mitigates these constraints by partitioning large circuits into smaller subcircuits, enabling execution on resource-limited devices. 
However, finding optimal circuit partitions remains a significant challenge, affecting both computational efficiency and accuracy.

To address these limitations, in this article, we propose CutQAS, a novel framework that integrates quantum circuit cutting with quantum architecture search (QAS) to enhance quantum chemistry simulations. 
Our RL-QAS framework employs a multi-step reinforcement learning (RL) agent to optimize circuit configurations. 
First, an RL agent explores all possible topologies to identify an optimal circuit structure. 
Subsequently, a second RL agent refines the selected topology by determining optimal circuit cuts, ensuring efficient execution on constrained hardware. 
Through numerical simulations, we demonstrate the effectiveness of our method in improving simulation accuracy and resource efficiency. 
This approach presents a scalable solution for quantum chemistry applications, offering a systematic pathway to overcoming hardware constraints in near-term quantum computing.
\end{abstract}


\section{Introduction}
Quantum computing is poised to be a promising paradigm for simulating quantum chemistry, offering the potential to outperform classical methods in modeling molecular physics and chemical reactions. 
Traditional approaches, such as quantum phase estimation (QPE) and variational quantum eigensolver (VQE), provide frameworks for solving electronic structure problems but are constrained by the limited qubit connectivity and coherence times of near-term quantum hardware \cite{cao2019quantum, ma2020quantum}.
Circuit cutting is a technique that partitions large quantum circuits into smaller subcircuits, allowing them to be executed on hardware with limited resources \cite{peng2020simulating}. 
This approach reduces quantum circuits' depth and qubit requirements, making them more feasible for near-term quantum processors. 
However, determining optimal circuit partitions remains a complex challenge that significantly impacts the efficiency and accuracy of simulations.
Quantum architecture search (QAS) is an emerging method that automates the design of quantum circuits \cite{sarkar2024automated} by optimizing their structure for specific tasks. 
By leveraging machine learning and heuristic algorithms, QAS identifies circuit configurations that balance accuracy, resource constraints, and noise resilience.

\begin{figure*}[ht]
    \centering 
    \includegraphics[clip, trim=0.0cm 0.0cm 0.0cm 6.2cm, width = 0.99\linewidth]{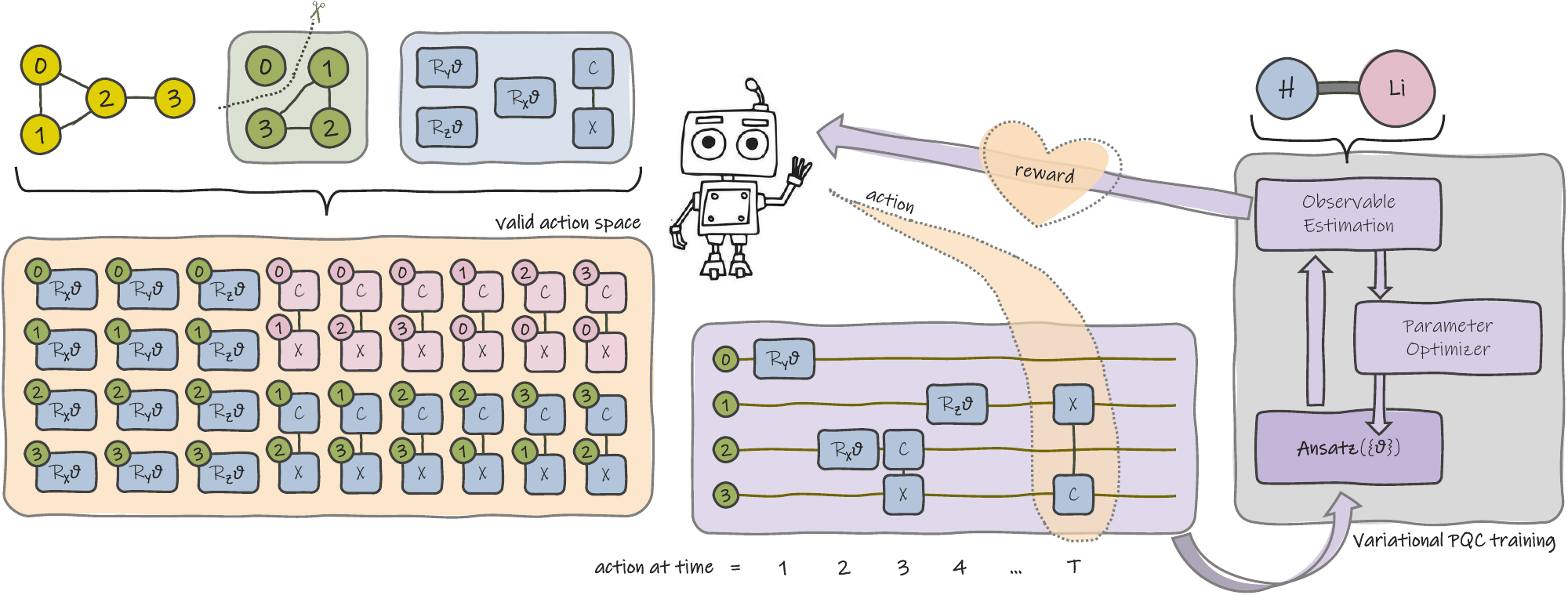}
    \caption{\textit{CutQAS workflow}: On a specific quantum processor topology (in yellow), a cut is considered (in green), and a gate set (in blue) is used; these generate a valid action space for the RL agent, which implements the CutQAS. A gate is added to the parametric quantum circuit at each action step. The PQC parameters are variationally optimized to solve the ground state energy of the selected molecule. The agent's reward reflects the estimate of the ground state energy (observable) obtained.}
    \label{fig:qcstack}
\end{figure*}

This work proposes a framework that combines quantum circuit cutting with quantum architecture search to enhance quantum chemistry simulations (see Fig.~\ref{fig:qcstack} for illustration). 
Integrating QAS with circuit cutting for quantum chemistry simulations presents a novel approach to tailoring quantum circuits to hardware limitations while maintaining computational accuracy.
By systematically exploring circuit architectures, we seek to mitigate hardware constraints and improve the feasibility of quantum simulations for complex molecular systems. 
We demonstrate our method’s effectiveness through numerical simulations and discuss its potential impact on the future of quantum chemistry.

\paragraph*{\textbf{Contributions}} We provide a unified RL-QAS framework to combine quantum circuit cutting with constrained topologies for variational quantum algorithms. To this end, we introduce a multi-step RL agent. In the first step, the agent searches for optimal topology by performing QAS on all possible topologies. In the subsequent step, the second agent takes the optimal topology from the previous agent and searches for an optimal cut on that topology by performing QAS for all possible qubit cuttings.  

The rest of the article in organized as follows.
In Section \ref{sec:background}, the techniques of quantum architecture search and quantum circuit cutting are introduced.
Section \ref{sec:cutqas} introduces the architecture and workflow of the proposed CutQAS method.
In Section \ref{sec:results}, the results for quantum chemistry simulations obtained via CutQAS are presented.
Section \ref{sec:conclusion} concludes the article with suggestions for future directions.

\section{Background} \label{sec:background}

In this section, we introduce the two techniques used in the design of the proposed CutQAS method: (i) quantum architecture search and (ii) quantum circuit cutting. 

\subsection{Quantum architecture search}

Quantum architecture search (QAS)~\cite{kuo2021quantum, ostaszewski2021reinforcement, fosel2021quantum} is a technique that automates the search for optimal quantum circuits for different information processing tasks. 
It comprises primarily of two parts. In the first part, a template of a parametrized quantum circuit is built. 
Following this, the circuit parameters are obtained via variational principle using a classical optimizer in a feedback loop. 
The algorithms constructed via this procedure are called Variational Quantum Algorithms (VQA). 
The parametrized circuit's design directly affects the solution's efficiency and expressivity and is a critical component of VQAs~\cite{cerezo2021variational}. 
In recent works~\cite{patel2024curriculum, kundu2024kanqas, sadhu2024quantum, he2024training}, QAS has been implemented inside the reinforcement learning (RL) framework where the variational circuit represents the RL state and given the state neural network structure is utilised to optimise the circuit structure and its parameters. 
QAS has also been used for designing quantum circuits as an approach to quantum program synthesis~\cite{kundu2024easy}. 
Apart from the RL-framework, simulated annealing~\cite{khatri2019quantum,zhou2020quantum,cincio2021machine}, unsupervised learning~\cite{sun2024quantum}, and training free framework~\cite{he2024training} has been also used for QAS. 

In ~\cite{patel2024curriculum, kuo2021quantum}, QAS under RL-framework has been utilised considering the constraint topologies of various quantum processors. 
However, an investigation of the potential of QAS under the circuit-cutting constraints has not been explored. 
In this work, we fill this gap by exploring the QAS under the RL framework when various topological constraints of QPUs and various cuts on the topologies are considered. We elaborate on this further in the following section.

\subsection{Quantum circuit cutting}

Quantum circuit cutting \cite{peng2020simulating} is a technique to extend the capabilities of quantum computation by decomposing large quantum circuits into smaller subcircuits that would fit the quantum volume of a noisy intermediate-scale quantum (NISQ) quantum processor.
Circuit cutting involves two primary types of cuts: gate cuts (space-like) and wire cuts (time-like) \cite{brenner2023optimal} and can be formally phrased in terms of a quasiprobability decomposition.
The smaller fragments are then executed separately, and the results are combined using classical post-processing to estimate the original circuit's outcome.
Generally, circuit cutting comes at the cost of classical post-processing overheads that are exponential in the number of cuts that are made to a circuit; however, these overheads might be avoided with suitably designed algorithms.
Circuit cutting can be conceptualized as performing tomography at the cut locations.

Recent advancements include efficient methods to reduce the classical and quantum resources required using techniques such as neglecting basis elements that pass no information \cite{chen2023efficient}, randomized measurements \cite{lowe2023fast}, sampling-based methods \cite{chen2022approximate}, and greedy-search~\cite{clark2023gtqcp} and special cases of circuit cutting~\cite{harrow2025optimal}. 
Circuit cutting can be beneficial in the reliability of the quantum computation even if a quantum device has enough qubits to 
simulate the entire circuit \cite{ayral2020quantum}.
Circuit cutting has also been experimentally demonstrated \cite{carrera2024combining} recently on two quantum processors of 127 qubits each.

\begin{figure*}[ht]
    \centering 
    \includegraphics[clip, trim=4.5cm 0.0cm 0.0cm 10.3cm, width = 0.85\linewidth]{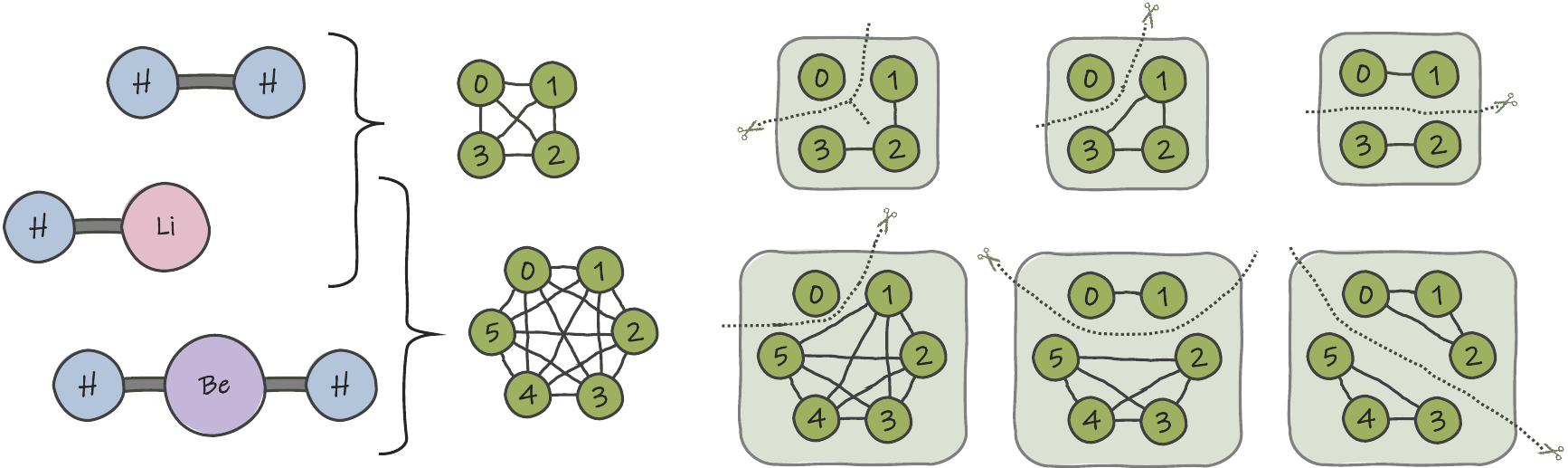}
    \caption{Set of experiments conducted for $3$ molecules and $6$ cut configurations on $4$ and $6$ qubits.}
    \label{fig:qcstacktopo}
\end{figure*}

\section{CutQAS Workflow} \label{sec:cutqas}

We introduce a reinforcement learning (RL)-based quantum circuit cutting under restricted quantum partition, namely CutQAS algorithm, illustrated in Fig..~\ref{fig:qcstack}. In the workflow of CutQAS, the RL agent interacts with the variational quantum framework to propose optimal topologies while obtaining the ground state energy of molecules with desired accuracies. The agent takes as input a \textit{3D tensor} (see Appendix~\ref{app:RLFramework} for more details) encoding of the structural details of the variational circuit. The 3D tensor encodes the circuit depth, positions of one-qubit and two-qubit gates, and the rotation parameters of parameterised gates. The set of gates that we consider in this work is $ \{R_x(\theta), R_y(\theta), R_z(\theta) \text{ and } CX\}$. Following this, the agent proposes a gate to add at the end of the circuit such that the variational minimum energy obtained trends towards the true ground state energy of the molecule. 

In our work, we consider a hybrid agent consisting of two sub-agents, namely \textit{agent-topo} and \textit{agent-cut} whose details are provided in Appendix~\ref{app:RLFramework} and the hyperparameters are provided in Table~\ref{tab:hyperparameter_list}. For a given molecule with a $k-$qubit Hamiltonian, the \textit{agent-topo} searches over the space of all possible topologies starting from a minimally connected topology to a $k-$connected topology. In each topology, the action (or the choice of the gate) of \textit{agent-topo} is restricted by the hardware connectivity between the physical qubits. For each search episode for each topology, \textit{agent-topo} adds a maximum of $n$ gates to the circuit and observes the deviation from chemical accuracy for the molecule. Then \textit{agent-topo} selects the topology with the best solution accuracy and circuit efficiency. \textit{agent-topo} then passes the best topology to \textit{agent-cut}.

\begin{figure}[ht]
    \centering 
    \includegraphics[clip, trim=24.0cm 0.0cm 0.0cm 8.4cm, width = 0.6\linewidth]{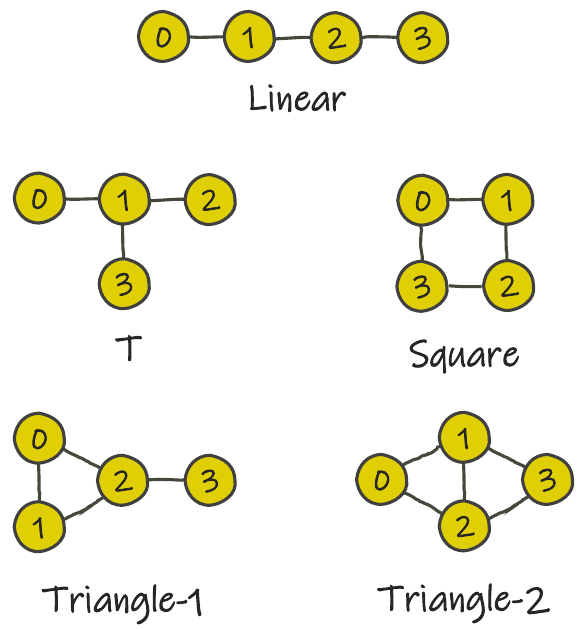}
    \caption{Quantum processor topologies considered.}
    \label{fig:topo}
\end{figure}

\textit{Agent-cut} on receiving the optimal topology from \textit{agent-cut} creates a list of all possible candidate actions. Such actions represent potential cut strategies while adhering to qubit physicality constraints. Intuitively speaking, each cut strategy corresponds to a specific way of partitioning the qubit space into sub-spaces. Then, following an approach similar to the action of \textit{agent-topo}, \textit{agent-cut} runs the optimization routine for all the possible cut strategies in parallel and selects the cut strategy that has optimal values of solution accuracy and circuit efficiency.

\begin{table}[]
    \centering
    \begin{tabular}{lllll}
        \toprule
         Mol & Basis & Mapping & Geometry & qubits \\  
         \midrule
         $\text{H}_2$& sto3g& Jordan-Wigner& H(0,0,0); H(0,0,0.7414) & 4 \\
         $\text{LiH}$ & sto3g& Parity& Li(0,0,0); H(0,0,3.4) & 4\\
         $\text{LiH}$ & sto3g& Jordan-Wigner& Li(0,0,0); H(0,0,3.4) & 6\\
         $\text{BeH}_2$ & sto3g& Jordan-Wigner& H(0,0,-1.33); Be(0,0,0); H(0,0,1.33)& 6\\
         \bottomrule
    \end{tabular}
    \caption{List of molecules in our simulation.}
    \label{tab:molecules_list}
\end{table}

\begin{table}[]
    \centering
    \begin{tabular}{llll}
        \toprule
    Batch size & $\text{H}_2$ & $\text{LiH}$ \\  
    \midrule
    Batch size & 1000 & 1000 \\
    Memory size & 20000 & 20000 \\
    Neurons & 1000 & 1000\\
    Hidden layers & 5 & 5\\
    Network optimizer & ADAM& ADAM\\
    Learning rate & 0.0001& 0.0001\\
    Number of steps & 40 & 70\\
    \bottomrule
    \end{tabular}
    \caption{List of hyperparameters.}
    \label{tab:hyperparameter_list}
\end{table}

\section{Results} \label{sec:results}

\subsection{Best topology selection}
In this stage, the \textit{agent-topo} operates on a 4-qubit $\texttt{H}_2$ molecule to identify the optimal topology. Subsequently, the \textit{agent-cut} takes the output from \textit{agent-topo}, which is the best topology, and applies all possible cuts to it. This process determines which cut is most suitable for the specific molecule.
\begin{table}[h!]
    \centering
    \caption{\small The \textit{agent-topo} generates the most compact parametrized quantum circuit to achieve the most accurate ground state solution when using Linear and Triangle-1 topologies. Given that the minimum error across all topologies is on the order of $10^{-8}$, we define the best topology as the one that achieves the highest accuracy in finding the ground state of the 4-qubit $\texttt{H}_2$ molecule with the fewest gates and smallest depth. Consequently, \textit{agent-topo} identifies Linear and Triangle-1 as the optimal topologies. In the subsequent step, \textit{agent-cut} will apply all possible cuts to these topologies.}
    \begin{tabular}{llllll}
        \toprule
        \textbf{Topology} & \textbf{Min Error ($\times 10^{-8}$)} & \textbf{Depth} & \textbf{CNOT} & \textbf{ROT} \\
        \midrule
        \cellcolor{lightgreen}Linear & \cellcolor{lightgreen}1.3072 & \cellcolor{lightgreen}7 & \cellcolor{lightgreen}6 & \cellcolor{lightgreen}4 \\
        Square & 1.3085 & 16 & 10 & 15 \\
        T & 1.3111 & 27 & 27 & 7 \\
        \cellcolor{lightgreen}Triangle-1 & \cellcolor{lightgreen}1.3112 & \cellcolor{lightgreen}7 & \cellcolor{lightgreen}7 & \cellcolor{lightgreen}1 \\
        Triangle-2 & 1.3071 & 9 & 7 & 4 \\
        \bottomrule
    \end{tabular}
    \label{tab:topo_agent_tab}
\end{table}

\paragraph{Definition of best topology} In the context of quantum architecture search~\cite{kundu2024enhancing,patel2024curriculum}, the best topology refers to the arrangement of quantum gates and qubits that achieves the most accurate solution for a specific problem (e.g., finding the ground state of a molecule) with the fewest resources. This typically involves minimizing circuit depth and the number of gates, particularly multi-qubit gates like CNOT, while maintaining a low error rate.

Our analysis in Tab.~\ref{tab:topo_agent_tab} demonstrates that the Linear and Triangle-1 topologies are optimal for achieving the ground state of the 4-qubit $\texttt{H}_2$ molecule, as they provide the best balance between accuracy and resource efficiency. Both topologies yield minimum errors on the order of $10^{-8}$, with the Linear topology using 6 CNOT gates and a depth of 7 and the Triangle-1 topology using 7 CNOT gates with the same depth. Notably, Triangle-1 requires fewer one-qubit rotations, making it particularly efficient in terms of gate count. In contrast, other topologies like Square and T require significantly more resources, with increased depths and gate counts. These findings highlight the importance of topology selection in quantum circuit optimization, underscoring the potential of Linear and Triangle-1 topologies for improving the efficiency of quantum simulations.

\subsection{Best cut selection}
Building on the results from the previous section, the \textit{agent-cut} applies possible cuts to the best topology identified by \textit{agent-topo}. For the 4-qubit $\texttt{H}_2$ problem, two types of cuts are considered: $1+3$, where one qubit is disconnected from the other three, and $2+2$, where the system is divided into two equal subsystems. The $1+3$ cut allows for crosstalk between qubits depending on the topology applied, and the best topology restrictions are applied to the three-qubit subsystem.

The results are summarized in Tables~\ref{tab:best_cut_4q_best} and~\ref{tab:best_cut_4q_avg}. Notably, the $2+2$ cut yields the worst approximation to the ground state of the $\texttt{H}_2$ molecule, while the $1+3$ cut performs significantly better. This suggests that a 3-qubit QPU can effectively solve the 4-qubit $\texttt{H}_2$ molecule. In Table~\ref{tab:best_cut_4q_best}, we observe that under the $1+3$ cut, both Linear and Triangle topologies achieve similar accuracy, but the Linear topology does so with fewer gates and less depth.

\begin{table}[h!]
    \centering
    \caption{Best across $5$ random initialisation of the neural network.}
    \begin{tabular}{l l l l l}
        \toprule
        \textbf{Cut \& Topology} & \textbf{Min Error} & \textbf{CX} & \textbf{One-Qubit Gates} & \textbf{Depth} \\
        \midrule
        \cellcolor{lightgreen}$1+3$, Linear & \cellcolor{lightgreen}$1.307\times 10^{-8}$ & \cellcolor{lightgreen}5 & \cellcolor{lightgreen}4 & \cellcolor{lightgreen}6 \\
        $1+3$, Triangle & $1.308\times 10^{-8}$ & 6 & 5 & 7 \\
        $2+2$ & $1.884\times 10^{-2}$ & 5 & 1 & 4 \\
        \bottomrule
    \end{tabular}
    \label{tab:best_cut_4q_best}
\end{table}

\begin{table}[h!]
    \centering
    \caption{Average across $5$ random initialisation of the neural network}
    \begin{tabular}{l l l l l}
        \toprule
        \textbf{Cut \& Topology} & \textbf{Avg. Error} & \textbf{CX} & \textbf{One-Qubit Gates} & \textbf{Depth} \\
        \midrule
        $1+3$, Linear & $1.311\times 10^{-8}$ & 10.2 & 2.8 & 10.0 \\
        \cellcolor{lightgreen}$1+3$, Triangle & \cellcolor{lightgreen}$1.310\times 10^{-8}$ & \cellcolor{lightgreen}7.0 & \cellcolor{lightgreen}2.4 & \cellcolor{lightgreen}6.8 \\
        $2+2$ & $1.884\times 10^{-2}$ & 5.4 & 1.4 & 5.4 \\
        \bottomrule
    \end{tabular}
    \label{tab:best_cut_4q_avg}
\end{table}

\begin{figure}
    \centering
    \includegraphics[width=\linewidth]{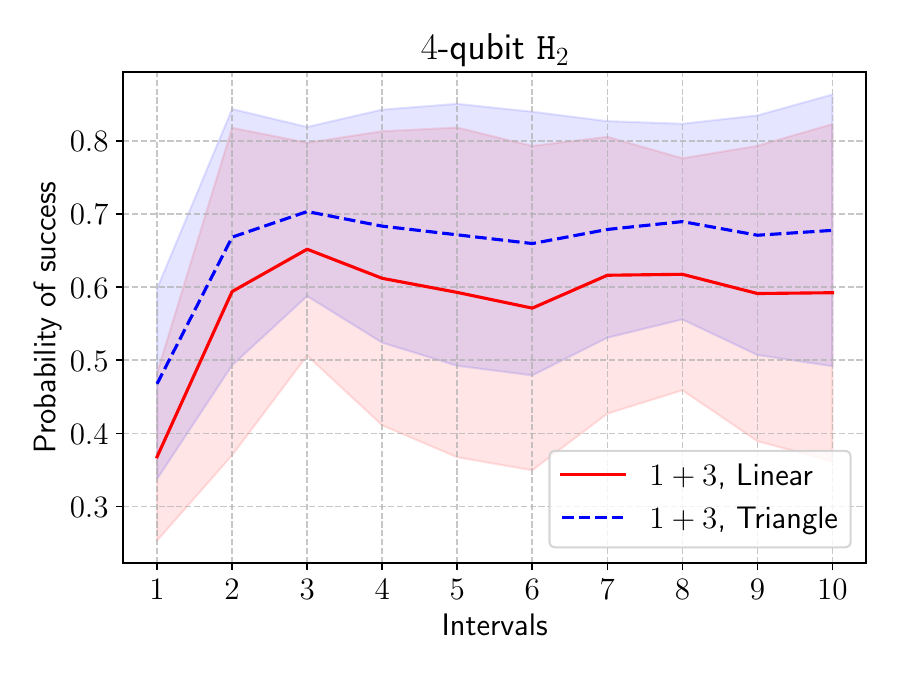}
    \caption{\small The probability of success with increasing intervals is higher with the \textit{Triangle} topology than the \textit{Linear} using the $1+3$ cut. The average is taken over $5$ different initializations of the neural network in the \textit{agent-cut}.}
    \label{fig:probability_of_success_4q}
\end{figure}

While the Linear topology excels in minimizing circuit depth and gate count, consistently achieving successful episodes proves more challenging compared to the Triangle topology shown in Fig.~\ref{fig:probability_of_success_4q}. This disparity arises from the inherent connectivity differences between the two topologies. All three qubits are interconnected in the Triangle topology, allowing for more flexible and robust quantum operations. This enhanced connectivity facilitates better crosstalk and interaction among qubits, which is crucial for maintaining a high probability of success across different episodes.

In contrast, the Linear topology, although efficient in terms of resource usage, has more restricted connectivity. Each qubit is connected only to its immediate neighbors, limiting the potential for complex interactions and crosstalk. This restricted connectivity can lead to a drop in the probability of success, as the system may not fully leverage the quantum advantages offered by more interconnected architectures.

\begin{figure}
    \centering
    \includegraphics[width=\linewidth]{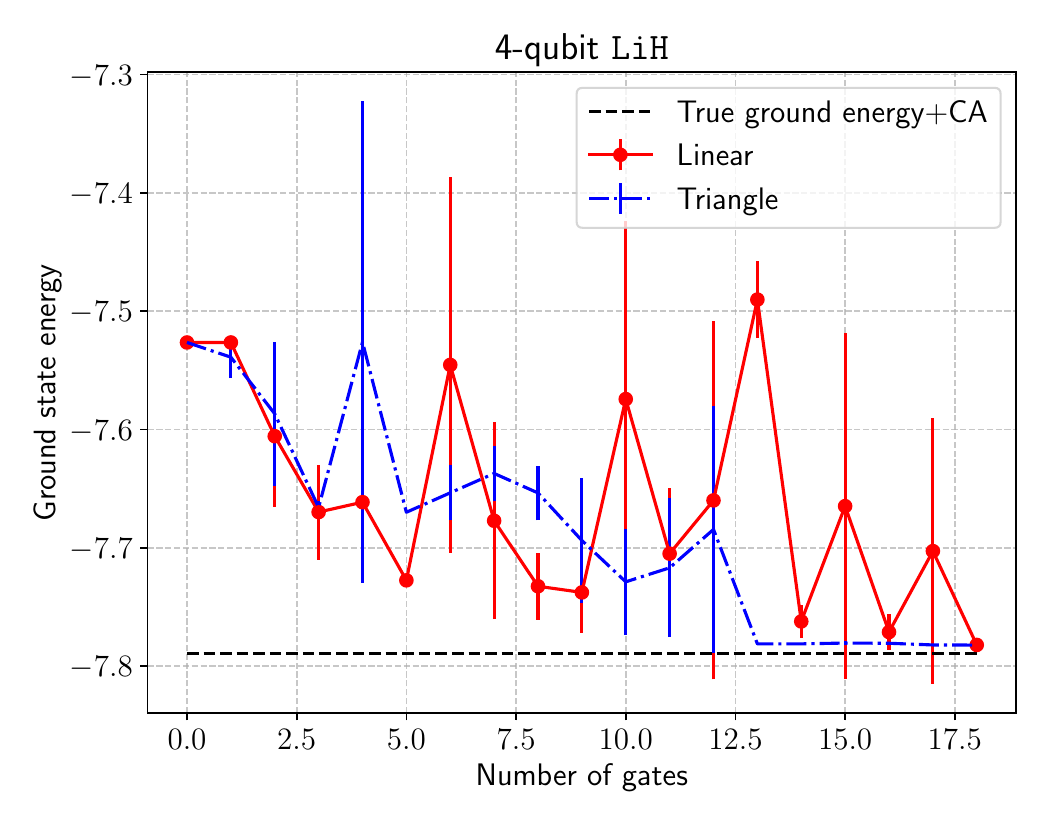}
    \caption{Results of achieved ground state energy for \texttt{LiH} on 4 qubits using CutQAS.}
    \label{fig:LiH_6_qubit}
\end{figure}

In Table~\ref{tab:best_cut_4q_avg} we compares different cuts for five different initialisations of the neural network. We observe that the $1+3$ Triangle topology has the minimum error with the lowest number of one-qubit and two-qubit gates on average over the different initialisations. This indicates that the $1+3$ Triangle setting is stable for solving the problem. We attribute this stability to the triangular topology being fully connected.

\begin{table}[h!]
    \centering
    \caption{$\texttt{BeH}_2$ molecule is more suitable for the RL-assisted quantum circuit cutting than $\texttt{LiH}$ providing lower error with smaller gates and circuit depth.}
    \begin{tabular}{l l l l l}
        \toprule
        \textbf{Molecule} & \textbf{Cut \& Topology} & \textbf{Min Error} & \textbf{Gates} & \textbf{Depth} \\
        \midrule
        $\texttt{LiH}$ (6q) & $2+4$ & $3.7\times 10^{-2}$ & 6 & 4 \\
        $\texttt{LiH}$ (6q) & $1+5$ & $2.6\times 10^{-2}$ & 59 & 36 \\
        \cellcolor{lightgreen}$\texttt{BeH}_2$ (6q) & \cellcolor{lightgreen}$3+3$ & \cellcolor{lightgreen}$5.9\times 10^{-3}$ & \cellcolor{lightgreen}5 & \cellcolor{lightgreen}4 \\
        \bottomrule
    \end{tabular}
    \label{tab:best_cut_6q_best}
\end{table}

In Table~\ref{tab:best_cut_6q_best}, we apply the RL-framework to find the ground state of 6-qubit $\texttt{LiH}$ and $\texttt{BeH}_2$ molecules under $1+5$, $2+4$ and $3+3$ qubit partition where in each partition we apply all-to-all qubit connectivity. We observe that $\texttt{LiH}$ with $1+5$ cut gives us lower error as compared to $2+4$ cut at the cost of additional gates and depth in the circuit. Whereas $\texttt{BeH}_2$ on $3+3$ cut provides us much lower error with smaller circuit than $\texttt{LiH}$, making $\texttt{BeH}_2$ much more suitable for our framework and in general for distributed quantum computing.

\section{Conclusion} \label{sec:conclusion}

In recent literature, quantum architecture search presents itself as a promising paradigm for solving quantum chemistry simulations by automating the design of quantum circuits. 
In the NISQ era, the qubits are noisy, and their number is limited. 
Hence, cutting the quantum circuits into smaller units and running the different units on the available hardware (in parallel or serially) promises to be an effective method for reducing the effect of hardware constraints. 

This work combines quantum circuit cutting with quantum architecture search to solve complex quantum chemistry problems. 
Such an approach tailors the problem's solution space to the search space of multiple independent quantum processing units while adhering to the hardware limitations and maintaining computational accuracy. 
Following this approach, we systematically explore the space of possible quantum architectures to improve the feasibility of quantum simulations for complex molecular systems and, at the same time, mitigate errors due to hardware constraints.  
    
We demonstrate the applicability of our approach by estimating the ground state of $4$-qubit $\text{H}_2$ and LiH molecules. Specifically, we introduce two search agents, agent-topo and agent-cut, which have different tasks. 
The agent-topo searches for an optimal solution in the space of all possible qubit processor topologies ranging from minimally connected topology to a fully connected topology. 
Once an optimal topology is found, the agent-cut considers the space of all possible cut topologies within the hardware constraints and finds the optimal cut topology for solving the problem. 

For the 4-qubit $\text{H}_2$ molecule, we observe that agent-topo identifies the linear and triangular topologies as optimal for achieving the ground state having a minimum error of $~1.31 \times 10^{-8}$ along with similar depth and CNOT counts. 
However, the linear topology requires more single qubit rotation gates. 
The other topologies (square and T) require significantly more resources. 

Next, when comparing cut strategies, we observe that the $1 + 3$ cut topology significantly outperforms the $2 + 2$ cut topology when approximating the ground state of molecular $\text{H}_2$. This indicates that the 4-qubit $\text{H}_2$ molecule can be effectively solved using a 3-qubit quantum processor. These observations indicate the importance of optimal topology selection when solving a class of problems in quantum circuit optimization. 

For future works, an interesting direction will be to analyse the connection between the ansatz obtained from the optimal cut topology and the inherent symmetries \cite{larocca2022group,meyer2023exploiting} of the problem Hamiltonian. 
This can potentially provide insights into the structure of the Hamiltonian and also into the sub-spaces of the Hamiltonian that are critical for solving the problem.

Another possible direction would be to allow the RL agent to implement a constant minimum number of CNOT gates between two non-local quantum processors over the quantum internet via the TeleGate method~\cite{situ2024distributed}. 
This will allow the utilisation of the full potential of non-local distributed quantum computing. An important and costly resource in such a protocol is the shared entanglement between the processors, which may be achieved via satellite links between the quantum computing stations~\cite{sadhu2023practical}. 
Hence, it will be worthwhile to analyse the dependence between the amount of shared entanglement and the structure of the solved Hamiltonian, a task we leave for future work. 

We anticipate that the methodology highlighted in this work will potentially enable computation of the ground state energies of large molecules by the distribution of the resource requirements across a full-stack quantum accelerator \cite{bertels2020quantum} and can have implications in the analysis of chemical reaction mechanisms, finding critical hidden sub-structures in large molecular Hamiltonians, and many others.


\section{Software availability} \label{sec:sw}
The open-sourced code for the project, configuration files, output data, and plotting codes for the
experiments presented in this article are available at: \href{https://github.com/Advanced-Research-Centre/CutQAS}{https://github.com/Advanced-Research-Centre/CutQAS}.

\section{Acknowledgement} \label{sec:ack}
A. Sadhu thanks the International Institute of Information Technology, Hyderabad, for supporting this project with computational and data storage resources during the employment. A.K. acknowledges funding from the Research Council of Finland through the Finnish Quantum Flagship project 358878 (UH).

\bibliographystyle{unsrt}
\bibliography{ref.bib}

\begin{thebibliography}{10}

\bibitem{cao2019quantum}
Yudong Cao, Jonathan Romero, Jonathan~P Olson, Matthias Degroote, Peter~D Johnson, M{\'a}ria Kieferov{\'a}, Ian~D Kivlichan, Tim Menke, Borja Peropadre, Nicolas~PD Sawaya, et~al.
\newblock Quantum chemistry in the age of quantum computing.
\newblock {\em Chemical reviews}, 119(19):10856--10915, 2019.

\bibitem{ma2020quantum}
He~Ma, Marco Govoni, and Giulia Galli.
\newblock Quantum simulations of materials on near-term quantum computers.
\newblock {\em npj Computational Materials}, 6(1):85, 2020.

\bibitem{peng2020simulating}
Tianyi Peng, Aram~W Harrow, Maris Ozols, and Xiaodi Wu.
\newblock Simulating large quantum circuits on a small quantum computer.
\newblock {\em Physical review letters}, 125(15):150504, 2020.

\bibitem{sarkar2024automated}
Aritra Sarkar.
\newblock Automated quantum software engineering.
\newblock {\em Automated Software Engineering}, 31(1):1--17, 2024.

\bibitem{kuo2021quantum}
En-Jui Kuo, Yao-Lung~L Fang, and Samuel Yen-Chi Chen.
\newblock Quantum architecture search via deep reinforcement learning.
\newblock {\em arXiv preprint arXiv:2104.07715}, 2021.

\bibitem{ostaszewski2021reinforcement}
Mateusz Ostaszewski, Lea~M Trenkwalder, Wojciech Masarczyk, Eleanor Scerri, and Vedran Dunjko.
\newblock Reinforcement learning for optimization of variational quantum circuit architectures.
\newblock {\em Advances in neural information processing systems}, 34:18182--18194, 2021.

\bibitem{fosel2021quantum}
Thomas F{\"o}sel, Murphy~Yuezhen Niu, Florian Marquardt, and Li~Li.
\newblock Quantum circuit optimization with deep reinforcement learning.
\newblock {\em arXiv preprint arXiv:2103.07585}, 2021.

\bibitem{cerezo2021variational}
Marco Cerezo, Andrew Arrasmith, Ryan Babbush, Simon~C Benjamin, Suguru Endo, Keisuke Fujii, Jarrod~R McClean, Kosuke Mitarai, Xiao Yuan, Lukasz Cincio, et~al.
\newblock Variational quantum algorithms.
\newblock {\em Nature Reviews Physics}, 3(9):625--644, 2021.

\bibitem{patel2024curriculum}
Yash~J. Patel, Akash Kundu, Mateusz Ostaszewski, Xavier Bonet-Monroig, Vedran Dunjko, and Onur Danaci.
\newblock Curriculum reinforcement learning for quantum architecture search under hardware errors.
\newblock In {\em The Twelfth International Conference on Learning Representations}, 2024.

\bibitem{kundu2024kanqas}
Akash Kundu, Aritra Sarkar, and Abhishek Sadhu.
\newblock Kanqas: Kolmogorov-arnold network for quantum architecture search.
\newblock {\em EPJ Quantum Technology}, 11(1):76, 2024.

\bibitem{sadhu2024quantum}
Abhishek Sadhu, Aritra Sarkar, and Akash Kundu.
\newblock A quantum information theoretic analysis of reinforcement learning-assisted quantum architecture search.
\newblock {\em Quantum Machine Intelligence}, 6(2):49, 2024.

\bibitem{he2024training}
Zhimin He, Maijie Deng, Shenggen Zheng, Lvzhou Li, and Haozhen Situ.
\newblock Training-free quantum architecture search.
\newblock In {\em Proceedings of the AAAI conference on artificial intelligence}, volume~38, pages 12430--12438, 2024.

\bibitem{kundu2024easy}
Akash Kundu and Leopoldo Sarra.
\newblock From easy to hard: Tackling quantum problems with learned gadgets for real hardware.
\newblock {\em arXiv preprint arXiv:2411.00230}, 2024.

\bibitem{khatri2019quantum}
Sumeet Khatri, Ryan LaRose, Alexander Poremba, Lukasz Cincio, Andrew~T Sornborger, and Patrick~J Coles.
\newblock Quantum-assisted quantum compiling.
\newblock {\em Quantum}, 3:140, 2019.

\bibitem{zhou2020quantum}
Xiangzhen Zhou, Sanjiang Li, and Yuan Feng.
\newblock Quantum circuit transformation based on simulated annealing and heuristic search.
\newblock {\em IEEE Transactions on Computer-Aided Design of Integrated Circuits and Systems}, 39(12):4683--4694, 2020.

\bibitem{cincio2021machine}
Lukasz Cincio, Kenneth Rudinger, Mohan Sarovar, and Patrick~J Coles.
\newblock Machine learning of noise-resilient quantum circuits.
\newblock {\em PRX Quantum}, 2(1):010324, 2021.

\bibitem{sun2024quantum}
Yize Sun, Zixin Wu, Yunpu Ma, and Volker Tresp.
\newblock Quantum architecture search with unsupervised representation learning.
\newblock {\em arXiv preprint arXiv:2401.11576}, 2024.

\bibitem{brenner2023optimal}
Lukas Brenner, Christophe Piveteau, and David Sutter.
\newblock Optimal wire cutting with classical communication.
\newblock {\em arXiv preprint arXiv:2302.03366}, 2023.

\bibitem{chen2023efficient}
Daniel~T Chen, Ethan~H Hansen, Xinpeng Li, Vinooth Kulkarni, Vipin Chaudhary, Bin Ren, Qiang Guan, Sanmukh Kuppannagari, Ji~Liu, and Shuai Xu.
\newblock Efficient quantum circuit cutting by neglecting basis elements.
\newblock In {\em 2023 IEEE International Parallel and Distributed Processing Symposium Workshops (IPDPSW)}, pages 517--523. IEEE, 2023.

\bibitem{lowe2023fast}
Angus Lowe, Matija Medvidovi{\'c}, Anthony Hayes, Lee~J O'Riordan, Thomas~R Bromley, Juan~Miguel Arrazola, and Nathan Killoran.
\newblock Fast quantum circuit cutting with randomized measurements.
\newblock {\em Quantum}, 7:934, 2023.

\bibitem{chen2022approximate}
Daniel Chen, Betis Baheri, Vipin Chaudhary, Qiang Guan, Ning Xie, and Shuai Xu.
\newblock Approximate quantum circuit reconstruction.
\newblock In {\em 2022 IEEE International Conference on Quantum Computing and Engineering (QCE)}, pages 509--515. IEEE, 2022.

\bibitem{clark2023gtqcp}
Joseph Clark, Travis~S Humble, and Himanshu Thapliyal.
\newblock Gtqcp: Greedy topology-aware quantum circuit partitioning.
\newblock In {\em 2023 IEEE International Conference on Quantum Computing and Engineering (QCE)}, volume~1, pages 739--744. IEEE, 2023.

\bibitem{harrow2025optimal}
Aram~W Harrow and Angus Lowe.
\newblock Optimal quantum circuit cuts with application to clustered hamiltonian simulation.
\newblock {\em PRX Quantum}, 6(1):010316, 2025.

\bibitem{ayral2020quantum}
Thomas Ayral, Fran{\c{c}}ois-Marie Le~R{\'e}gent, Zain Saleem, Yuri Alexeev, and Martin Suchara.
\newblock Quantum divide and compute: Hardware demonstrations and noisy simulations.
\newblock In {\em 2020 IEEE Computer Society Annual Symposium on VLSI (ISVLSI)}, pages 138--140. IEEE, 2020.

\bibitem{carrera2024combining}
Almudena Carrera~Vazquez, Caroline Tornow, Diego Rist{\`e}, Stefan Woerner, Maika Takita, and Daniel~J Egger.
\newblock Combining quantum processors with real-time classical communication.
\newblock {\em Nature}, pages 1--5, 2024.

\bibitem{kundu2024enhancing}
Akash Kundu, Przemys{\l}aw Bede{\l}ek, Mateusz Ostaszewski, Onur Danaci, Yash~J Patel, Vedran Dunjko, and Jaros{\l}aw~A Miszczak.
\newblock Enhancing variational quantum state diagonalization using reinforcement learning techniques.
\newblock {\em New Journal of Physics}, 26(1):013034, 2024.

\bibitem{larocca2022group}
Mart{\'\i}n Larocca, Fr{\'e}d{\'e}ric Sauvage, Faris~M Sbahi, Guillaume Verdon, Patrick~J Coles, and Marco Cerezo.
\newblock Group-invariant quantum machine learning.
\newblock {\em PRX quantum}, 3(3):030341, 2022.

\bibitem{meyer2023exploiting}
Johannes~Jakob Meyer, Marian Mularski, Elies Gil-Fuster, Antonio~Anna Mele, Francesco Arzani, Alissa Wilms, and Jens Eisert.
\newblock Exploiting symmetry in variational quantum machine learning.
\newblock {\em PRX quantum}, 4(1):010328, 2023.

\bibitem{situ2024distributed}
Haozhen Situ, Zhimin He, Shenggen Zheng, and Lvzhou Li.
\newblock Distributed quantum architecture search.
\newblock {\em Physical Review A}, 110(2):022403, 2024.

\bibitem{sadhu2023practical}
Abhishek Sadhu, Meghana~Ayyala Somayajula, Karol Horodecki, and Siddhartha Das.
\newblock Practical limitations on robustness and scalability of quantum internet.
\newblock {\em arXiv preprint arXiv:2308.12739}, 2023.

\bibitem{bertels2020quantum}
Koen Bertels, Aritra Sarkar, Thomas Hubregtsen, M~Serrao, Abid~A Mouedenne, Amitabh Yadav, A~Krol, Imran Ashraf, and C~Garcia Almudever.
\newblock Quantum computer architecture toward full-stack quantum accelerators.
\newblock {\em IEEE Transactions on Quantum Engineering}, 1:1--17, 2020.

\bibitem{peruzzo2014variational}
Alberto Peruzzo, Jarrod McClean, Peter Shadbolt, Man-Hong Yung, Xiao-Qi Zhou, Peter~J Love, Al{\'a}n Aspuru-Guzik, and Jeremy~L O’brien.
\newblock A variational eigenvalue solver on a photonic quantum processor.
\newblock {\em Nature communications}, 5(1):4213, 2014.

\bibitem{tilly2022variational}
Jules Tilly, Hongxiang Chen, Shuxiang Cao, Dario Picozzi, Kanav Setia, Ying Li, Edward Grant, Leonard Wossnig, Ivan Rungger, George~H Booth, et~al.
\newblock The variational quantum eigensolver: a review of methods and best practices.
\newblock {\em Physics Reports}, 986:1--128, 2022.

\bibitem{melo2001convergence}
Francisco~S Melo.
\newblock Convergence of q-learning: A simple proof.
\newblock {\em Institute Of Systems and Robotics, Tech. Rep}, pages 1--4, 2001.

\end{thebibliography}

\vspace{12pt}

\appendix

\subsection{RL framework} \label{app:RLFramework}

The RL framework in CutQAS follows a feedback-driven curriculum learning method introduced in~\cite{ostaszewski2021reinforcement} where the RL-state is encoded using the tensor-based one-hot encoding method described in~\cite{patel2024curriculum}. However, the tensor's dimensions vary depending on the size of the action space. The RL-state encoding translates a quantum circuit into a tensor of
\begin{equation}
    \text{size} = D_\text{max}\times N\times(N+N_\text{1q})\label{eq:tensor_based_encoding_eqn},
\end{equation}
where $D_\text{max}$ is a hyperparameter and is defined as the maximum allowed gates per episode, i.e., the length of an episode, $N$ is the number of qubits, and $N_\text{1q}$ defines the number of 1-qubit gates. The first $N\times N$ encodes the position of the 2-qubit gate, and the remaining $N\times N_\text{1q}$ encodes the position of the 1-qubit gate. 

We initialize the RL state with an empty quantum circuit. Utilizing the following reward function 
\begin{equation} \label{eq:reward}
\small
R= \begin{cases}5 & \text { if } C_t<\xi, \\ 
-5 & \text { if } t \geq D_\text{max} \text { and } C_t \geq \xi, \\ 
\max \left(\frac{C_{t-1} - C_t}{C_{t-1} - C_{\min }},-1\right) & \text { otherwise }\end{cases}
\end{equation}
the RL agent decides on the next action through an $\epsilon$-greedy policy. In the reward function, $C_t$ is the cost function (in our case VQE energy, see Eq.~\ref{eq:vqe_cost}) at step $t$ and $\xi$ is a hyperparameter (for VQE it is the chemical accuracy $0.0016$ Hartree).
The action is chosen from a predefined action space $(\mathbb{A})$ which contains parametrized $1$- and non-parameterized $2$-qubit gates i.e. $\mathbb{A} = \{CX, RX, RY, RZ\}$.
Depending on the action, the RL-state is modified in the next step $t+1$. 

In this framework, we implement the variational quantum eigensolver (VQE)~\cite{peruzzo2014variational,tilly2022variational} to find the ground state of $\texttt{H}_2$ and $\texttt{LiH}$ molecules under different topologies of quantum processors and different ways to partition the processor. In VQE, the objective is to find the ground state energy of a chemical Hamiltonian $H$ by minimizing the energy 
\begin{equation}
    C(\vec{\theta}) = \min_{\Vec{\theta}}\left(\langle\psi(\Vec{\theta})|H|\psi(\Vec{\theta})\rangle \right).
    \label{eq:vqe_cost}
\end{equation}
The trial state $|\psi(\Vec{\theta})\rangle$ is prepared by applying a parameterized quantum circuit (PQC), $U(\vec{\theta})$, to the initial state $|\psi_{\text{initial}}\rangle$, where $\vec{\theta}$ specify the rotation angles of the local unitary operators in the circuit.

\paragraph{Hyperparameters:} We set the discount factor ($\gamma$) to 0.88. We implemented an $\epsilon$-greedy policy for selecting random actions, with $\epsilon$ decaying by a factor of 0.99995 per step from an initial value of $\epsilon = 1$ until it reached a minimum value of $\epsilon = 0.05$. The memory replay buffer size was fixed at 20000, and the target network in the DQN training process was updated after every 500 actions. We implemented a testing phase in the RL framework after every 100 training episodes. In this testing phase, we set the randomness factor to $\epsilon = 0$ to halt the random exploration and ensure a set of deterministic actions. We exclude the experiences acquired in this phase from the memory replay buffer. We greedily adjusted the threshold after $G = 500$ episodes for both noiseless and noisy 3- and 4-qubit problems with an amortization radius set at $\delta = 0.0001$. This amortization radius decreased by $\delta/\kappa = 0.00001$ after every 50 successfully solved episodes, beginning from an initial threshold value of $0.005$.

\subsection{Double deep Q-network}\label{appndx:ddqn}
Deep reinforcement learning techniques utilize neural networks (NN) to adapt the agent's policy to optimize the return:
\begin{eqnarray}
G_t = \sum_{k=0}^{\infty} \gamma^k r_{t+k+1},
\end{eqnarray}
where $\gamma \in [0,1)$ is the discount factor. For each state-action pair $(s,a)$, a value is assigned, quantifying the expected return at step $t$ under policy $\pi$:
\begin{eqnarray}
q_\pi (s,a) = \mathbbm{E}_\pi [G_t | s_t = s, a_t = a].
\end{eqnarray}

The goal is to determine the optimal policy that maximizes the expected return, which can be derived from the optimal action-value function $q_\ast$, defined by the Bellman optimality equation:
\begin{eqnarray}
q_\ast (s,a) = \mathbbm{E} \bigg[ r_{t+1} + \max_{a'} q_\ast (s_{t+1},a') | s_t = s, a_t = a \bigg].
\end{eqnarray}

Instead of solving the Bellman optimality equation, value-based RL learns the optimal action-value function from data samples. 
Q-learning is a prominent value-based RL algorithm, where each state-action pair $(s,a)$ is assigned a Q-value $Q(s,a)$, which is updated to approximate $q_\ast$. Starting from randomly initialized values, the Q-values are updated as:
\begin{eqnarray}
&&Q(s_t,a_t) \gets Q(s_t,a_t) \nonumber \\
&& \qquad \qquad + \alpha \bigg( r_{t+1} + \gamma \max_{a'} Q(s_{t+1},a') - Q(s_t,a_t) \bigg),
\nonumber \\
\end{eqnarray}
where $\alpha$ is the learning rate, $r_{t+1}$ is the reward at time $t+1$, and $s_{t+1}$ is the state encountered after taking action $a_t$ in state $s_t$. This update rule is proven to converge to the optimal Q-values in the tabular case if all $(s,a)$ pairs are visited infinitely often~\cite{melo2001convergence}. To ensure sufficient exploration in a Q-learning setting, an $\epsilon$-greedy policy is used, defined as:
\begin{align}
\pi(a|s) := 
\begin{cases}
1-\epsilon_t & \text{if $a = \max_{a'} Q(s,a')$},\\
\epsilon_t & \text{otherwise}.
\end{cases}
\end{align}

The $\epsilon$-greedy policy introduces randomness to the actions during training. After training, a deterministic policy is used.
    
NN and function approximations are employed to extend Q-learning to large state and action spaces. 
The NN training typically requires independently and identically distributed data. This problem is circumvented by experience replay. This method divides experiences into single-episode updates and creates batches that are randomly sampled from memory. For stabilizing training, two NNs are used: a policy network, which is continuously updated, and a target network, which is an earlier copy of the policy network. The policy network estimates the current value, while the target network provides a more stable target value $Y$ given by :
\begin{eqnarray}
Y_{\text{DQN}} = r_{t+1} + \gamma \max_{a'} Q_{\text{target}} (s_{t+1},a').
\end{eqnarray}

In the double deep Q-network (DDQN) algorithm, we sample the action for the target value from the policy network to reduce the overestimation bias present in standard DQN. The corresponding target is defined as:
\begin{eqnarray}
Y_{\text{DDQN}} = r_{t+1} + \gamma Q_{\text{target}} \bigg( s_{t+1}, \arg \max_{a'} Q_{\text{policy}} (s_{t+1},a') \bigg).
\end{eqnarray}
This target value is approximated via a loss function. In this work, we consider the loss function as the smooth L1-norm.

\end{document}